\documentstyle{article}

\newcommand{\eps}{\varepsilon}

\newtheorem{proposition}{Proposition}[section]
\newtheorem{theorem}[proposition]{Theorem}

\newtheorem{lemma}[proposition]{Lemma}

\newenvironment{proof}{\noindent{\bf Proof:$\;$}}{\hfill$\Box$\medskip}

\newcommand{\art}[6]{{\rm #1, \rm #2, \it #3 \bf #4 \rm (#5), \mbox{#6}.}}
\newcommand{\artnopage}[5]{{\rm #1, \rm #2, \it #3 \bf #4 \rm (#5).}}

\newcommand{\book}[3]{{\rm #1, \it #2, \rm #3.}}

\def\ring#1{\stackrel{\circ}{#1}}

\begin{document}

\centerline{\Large{\bf LOCAL EXISTENCE OF SPINOR-
}}

\centerline{\Large{\bf AND TENSOR POTENTIALS 
}}

\

\centerline{{\bf F. Andersson} and {\bf S. B. Edgar}}

\centerline{ \it Department of Mathematics,
 Link\"oping University,}

\centerline{\it S581 83 Link\"oping,}

\centerline{\it Sweden. }

\centerline{Email: frand@mai.liu.se, bredg@mai.liu.se}

\

\centerline{Short title: Spinor- and tensor potentials}

\

\centerline{{\bf Abstract}}

\

We give new simple direct proofs in all spacetimes for the existence 
of asymmetric $(n,m+1)$-spinor potentials for completely symmetric 
$(n+1,m)$-spinors and for the existence of symmetric $(n,1)$-spinor
potentials for symmetric $(n+1,0)$-spinors. These proofs introduce a
`superpotential', i.e., a potential of the potential, which also enables
us to get explicit statements of the gauge freedom of the original
potentials. The main application for these results is the Lanczos potential
$L_{ABCA'}$, of the Weyl spinor and the electromagnetic vector potential
$A_{AA'}$. We also investigate the possibility of existence of a 
{\em symmetric} potential $H_{ABA'B'}$ for the Lanczos potential, and prove
that in {\em all Einstein spacetimes} any symmetric (3,1)-spinor $L_{ABCA'}$
possesses a symmetric potential $H_{ABA'B'}$. Potentials of this type have
been found earlier in investigations of some very special spinors in
restricted classes of spacetimes. All of the new spinor results are
translated into tensor notation, and where possible given also for four
dimensional spaces of arbitrary signature.

\

\section{Introduction}

More than 30 years ago Lanczos \cite{Lanczos} proposed a first order
potential for the Weyl tensor. However, in 1983 Bampi and Caviglia 
\cite{BC} showed that Lanczos' original proof was flawed and supplied a
rigorous but complicated proof of local existence for four dimensional
analytic spaces, independent of signature. Illge \cite{Illge} has supplied
a more conventional proof of existence (by means of a Cauchy problem) in
spinor notation that, in its full generality, does not seem to generalize
in an obvious manner, to arbitrary signature. Moreover, it should be emphasized
that Illge's work has highlighted the simple and natural structure of the
Lanczos potential {\em in spinor notation}, and makes it clear that for work
in spacetimes ($C^{\infty}$ manifolds with Lorentz signature) the spinor
formalism is much simpler than the tensor formalism. It should also be noted
that in Lorentz signature the Lanczos potential satisfies a wave equation, and
the well posedness of the corresponding Cauchy problem enabled Illge to remove
the assumption about analyticity in his proof.

It is important to note that the two existence proofs supplied by Bampi 
and Caviglia \cite{BC}, and by Illge \cite{Illge}, respectively, do not
directly concern the Weyl tensor/spinor $C_{abcd}/\Psi_{ABCD}$, but are
valid for {\em any} tensor/spinor $W_{abcd}/W_{ABCD}$ having the same
algebraic symmetries as the Weyl tensor/spinor. Furthermore, Illge's work
also discusses the existence of potentials for completely symmetric 
spinors with an arbitrary number of primed and unprimed indices; in
general these potentials are not symmetric.

The results in this paper for four dimensional spacetimes are given in 
spinor notation following the conventions of \cite{PR1} where the results
are natural and the calculations comparatively simple; however, we also give
the results in tensor notation for four dimensional spacetimes, and where
possible in four dimensional spaces of any signature. We remark that in
general the Lanczos potential does not exist in dimensions higher than four 
\cite{EH2}.

Note that a spinor $S_{A_{1}\cdots A_{n}B_{1}'\cdots B_{m}'}$ having 
both primed and unprimed indices is said to be {\em (completely) 
symmetric} if it is symmetric over both types of indices i.e.,
$$
 S_{A_{1}\cdots A_{n}B_{1}'\cdots B_{m}'}=
 S_{(A_{1}\cdots A_{n})(B_{1}'\cdots B_{m}')}
$$

In Section 2 we state Illge's theorem for the existence and uniqueness 
of a symmetric spinor potential $L_{A_{1}\cdots A_{n}A'}$ for the 
symmetric spinor $W_{AA_{1}\cdots A_{n}}$. We also state the 
analogous result for a spinor potential $L_{A_{1}\cdots 
A_{n}B_{1}'\cdots B_{m}'A'}(=L_{(A_{1}\cdots A_{n})(B_{1}'\cdots
B_{m}')A'}$ for the completely symmetric spinor $W_{AA_{1}\cdots
A_{n}B_{1}'\cdots B_{m}'}$. In addition we quote the corresponding Lanczos
wave equation for $L_{ABCA'}$ and $L_{A_{1}\cdots A_{n}B_{1}'\cdots
B_{m}'A'}$ showing how in the latter case an {\em algebraic} constraint
arises in general, if we try to demand a {\em completely symmetric}
$L_{A_{1}\cdots A_{n}B_{1}'\cdots B_{m}'A'}$

In Section 3 we give a new simple proof of local existence of a symmetric
Lanczos potential $L_{ABCA'}$ of an arbitrary symmetric spinor $W_{ABCD}$
in an arbitrary spacetime. An interesting aspect of the proof is that it
also involves a potential $T_{ABCD}=T_{(ABC)D}$ of $L_{ABCA'}$ which may
be important in itself. This proof also generalizes to spinors with other
index configurations.

Furthermore, it is straightforward to translate the existence proof 
for the Lanczos potential into tensor notation and adapt it into an existence 
proof for four dimensional analytic spaces of arbitrary signature. 

In Section 4 we examine the gauge freedom of the Lanczos potential. We
obtain an explicit formula for the gauge freedom involving the potential
$T_{ABCD}$, analogously to electromagnetic theory where it is known that
the gauge freedom in the electromagnetic potential (after its curl and
divergence are specified) is given by the gradient of a scalar field that
satisfies a certain wave equation. Again, the tensor version of this 
result is given.

As noted above, Illge has shown the existence of (asymmetric) potentials
for completely symmetric spinors with an arbitrary number of primed and
unprimed indices. Thus, a Lanczos potential $L_{ABCA'}$ of some symmetric
spinor $W_{ABCD}$ itself has spinor potentials. One example is the spinor
$T_{ABCD}$ referred to above, but there are reasons why we are more interested
in having a {\em symmetric} potential of the type $H_{ABA'B'}=H_{(AB)(A'B')}$
(see e.g., \cite{lic}, \cite{AE}, \cite{Bergqvist} and \cite{TdC1}). Although
such a potential does not exist in all spacetimes we demonstrate in Section 5
that it does exist in all Einstein spacetimes. In order to obtain a unique
solution to the problem we will supplement the defining equation for
$H_{ABA'B'}$ with certain other conditions and use a technique which is similar
in structure to Illge's proof for the existence of $L_{ABCA'}$. As a result our
proof of this result will lack the simplicity of the existence proof for
$L_{ABCA'}$ given in Section 3. A tensor version for spacetimes and 
four dimensional spaces of other signatures is also given.

In Section 6 we will look in more detail at the important application 
to electromagnetism in curved space. We do this in order to see how 
the results in the first sections relate to more familiar results on 
potentials such as Poincare's lemma, and what simplifications can be 
achieved due to the simpler index structure of the electromagnetic 
spinor, and due to Maxwell's equations.

In Section 7 we discuss how the results in this paper links up with existing
results and applications.

\section{Preliminaries}

Let $M$ be a spacetime (i.e., a real, $C^{\infty}$, 4-dimensional 
manifold with a metric of signature $(+---)$). For simplicity we will 
restrict ourselves to tensor- and spinor fields of class $C^{\infty}$,
but note that the results given could be generalized to tensor- and 
spinor fields of lesser regularity by using theorems on hyperbolic
systems where the fields are only assumed to be in some Sobolev space,
instead of the theorems used here. For definitions of the Levi-Civita
connection, the curvature spinors etc., we will follow the conventions
in \cite{PR1}. Also note that all indices (both tensor- and spinor 
indices) occurring in this paper are abstract indices \cite{PR1}.

Illge has shown \cite{Illge} that given any symmetric spinors $W_{ABCD}=
W_{(ABCD)}$, $F_{BC}=F_{(BC)}$ there exists (locally) a symmetric spinor
$L_{ABCA'}=L_{(ABC)A'}$ such that
$$
 W_{ABCD}=2\nabla_{(A}{}^{A'}L_{BCD)A'}\;\;,\;\;F_{BC}=\nabla^{AA'}
 L_{ABCA'}.
$$
The first of these equations is called the Weyl-Lanczos equation and
such a spinor $L_{ABCA'}$ is said to be a Lanczos (spinor) potential of 
$W_{ABCD}$. The spinor $F_{BC}$ is called the differential gauge of 
$L_{ABCA'}$. When $F_{BC}=0$ the Lanczos potential is said to be in 
Lanczos differential gauge. Of particular interest is the case $W_{ABCD}
=\Psi_{ABCD}$ i.e., Lanczos potentials of the Weyl curvature spinor.
These Lanczos potentials are spinor analogues of the Lanczos {\it tensor}
potentials, originally investigated in \cite{Lanczos}. For an extensive
account of the Lanczos potential and its properties, see \cite{lic} 
and \cite{EH}. 

One of the most remarkable results concerning Lanczos potentials is Illge's
wave equation \cite{Illge}. Suppose $L_{ABCA'}$ is a Lanczos potential of 
$W_{ABCD}$ in the differential gauge $F_{BC}$. Then $L_{ABCA'}$ satisfies
the following linear wave equation
$$
 \Box L_{ABCA'}+6\Phi_{A'B'(A}{}^{D}L_{BC)D}{}^{B'}+6\Lambda L_{ABCA'}
 +\nabla_{A'}^{D}W_{ABCD}-\frac{3}{2}\nabla_{A'(A}F_{BC)}=0
$$
Now, if $W_{ABCD}$ is actually the Weyl spinor $\Psi_{ABCD}$, if the 
spacetime is vacuum and if $L_{ABCA'}$ is in Lanczos differential gauge, 
we obtain the remarkably simple equation
$$
 \Box L_{ABCA'}=0.
$$

By letting $L_{abc}$ be the tensor equivalent of the hermitian spinor
$$
 L_{abc}=L_{ABCC'}\eps_{A'B'}+\overline{L}_{A'B'C'C}\eps_{AB},
$$
the tensor $L_{abc}$ has the symmetries $L_{abc}=L_{[ab]c}$, $L_{[abc]}=0$,
$L_{ab}{}^{b}=0$. This last symmetry was originally thought of as a gauge
condition called the Lanczos algebraic gauge; however, because of the spinor
correspondence we choose to include this symmetry in the definition of the
Lanczos potential. As we shall see below, it also gives us a 
comparatively simple form of the tensor equation corresponding to the 
Weyl-Lanczos equation.
 
We can now define a Lanczos {\em tensor} potential of the Weyl tensor
$\,C_{abcd}$, or indeed of any tensor $W_{abcd}$ having the same algebraic
symmetries as the Weyl tensor, by translating the Weyl-Lanczos equation
into tensor formalism. We obtain the Weyl-Lanczos {\em tensor} equation
which reads
\begin{equation}
 W_{abcd}=L_{ab[c;d]}+L_{cd[a;b]}-{}^{*}L^{*}_{ab[c;d]}-{}^{*}
 L^{*}_{cd[a;b]} \label{WLt}
\end{equation}
where $W_{abcd}$ has the same algebraic symmetries as the Weyl tensor.
This is the original definition of the Lanczos potential given in 
\cite{Lanczos}. By differentiating the Weyl-Lanczos tensor equation and
using the Bianchi identities and the commutators we obtain a wave equation, 
similar to Illge's spinor wave equation. It is
$$
 \Box L_{abc}-2L^{def}g_{c[a}C_{b]def}+2L_{[a}{}^{de}C_{b]edc}+
 \frac{1}{2}L^{de}{}_{c}C_{deab}=0
$$
in vacuum, Lanczos differential gauge and $W_{abcd}=C_{abcd}$.
It is interesting to note that it is much more difficult to calculate 
this tensor wave equation than the corresponding spinor one, and it 
also appears at first to have a much more complicated structure.
It includes an expression involving products of the Weyl tensor and the 
Lanczos potential explicitly. However, it has been confirmed that this 
additional expression is actually identically zero in four, and only 
in four dimensions \cite{EH} giving
$$
 \Box L_{abc}=0
$$
in agreement with the spinor equation. This is a consequence of Lovelock's 
identity \cite{Lovelock}. Therefore the Lanczos wave equation provides a 
striking illustration of the power of the spinor formalism.

An interesting result regarding this wave equation was proved by Edgar and
H\"oglund \cite{EH}. They showed that in a vacuum spacetime of 
`sufficient generality' (see \cite{EH} or \cite{BS}) a spinor $L_{ABCA'}$ in
Lanczos differential gauge $\nabla^{AA'}L_{ABCA'}=0$ is a constant 
multiple of a Lanczos potential of the Weyl spinor if and only if
$\Box L_{ABCA'}=0$. Hence, in this particular case Illge's wave equation is
actually a sufficient condition for $L_{ABCA'}$ to be a constant 
multiple of a Lanczos potential of the Weyl spinor.

Illge's theorem in \cite{Illge} is actually more general than we have quoted 
above. Illge proves the existence of a potential similar to the one 
mentioned above, for the case when the symmetric spinor $W$ has an 
arbitrary number of indices. For easy reference we include the complete
theorem of Illge in this section, together with a generalization
also mentioned in \cite{Illge}:
\begin{theorem}\label{Illge1}
 Let symmetric spinor fields $W_{AA_{1}\cdots A_{n}}$, $F_{A_{2}\cdots 
 A_{n}}$, a spacelike past-compact hypersurface $\Sigma$ of class $C^{\infty}$
 and a symmetric spinor field $\ring{L}_{A_{1}\cdots A_{n}A'}$ defined 
 only on $\Sigma$\footnote{From now on, a circle above a spinor field
 i.e., $\ring{L}$ will always mean that the spinor field is defined only on
 $\Sigma$.} be given. Then there exists a neighbourhood of $\Sigma$ in which
 the equations
 \begin{eqnarray}
  W_{AA_{1}\cdots A_{n}} & = & 2\nabla_{(A}{}^{A'}L_{A_{1}\cdots A_{n})A'}
  \nonumber \\ F_{A_{2}\cdots A_{n}} & = & \nabla^{A_{1}A'}L_{A_{1}A_{2}
  \cdots A_{n}A'}\nonumber.
 \end{eqnarray}
 have a unique symmetric solution $L_{A_{1}\cdots A_{n}A'}$ satisfying
 $L|_{\Sigma}=\ring{L}$. 
\end{theorem}
We note that a spinor $W_{ABCD}$ in general has many Lanczos potentials in
each differential gauge $F_{BC}$.

Following Illge \cite{Illge}Êwe attempt to generalize this theorem to symmetric
spinors with both primed and unprimed indices. Let $W_{AA_{1}\cdots A_{n}B_{1}'
\cdots B_{m}'}$ and $F_{A_{2}\cdots A_{n}B_{1}'\cdots B_{m}'}$ be completely
symmetric spinors. We then look for a spinor $L_{A_{1}\cdots A_{n}B_{1}'\cdots
B_{m}'A'}$ so that
\begin{eqnarray}
  W_{AA_{1}\cdots A_{n}B_{1}'\cdots B_{m}'} & = & 2\nabla_{(A}{}^{A'}L_{A_{1}
  \cdots A_{n})B_{1}'\cdots B_{m}'A'}\nonumber \\ F_{A_{2}\cdots A_{n}B_{1}'
  \cdots B_{m}'} & = & \nabla^{A_{1}A'}L_{A_{1}A_{2}\cdots A_{n}B_{1}'\cdots
  B_{m}'A'}\nonumber.
\end{eqnarray}
From this equation we see that it is natural to require that $L$ has 
the symmetry
$$
 L_{A_{1}\cdots A_{n}B_{1}'\cdots B_{m}'A'}=L_{(A_{1}\cdots A_{n})(B_{1}'\cdots
 B_{m}')A'}
$$
By combining the above two equations into one, differentiating and using
the commutators we arrive at a wave equation analogous to Illge's wave 
equation (these calculations will be shown in detail for some special 
cases in later sections)
\begin{eqnarray}
	0 & = & \Box L_{A_{1}\cdots A_{n}B_{1}'\cdots B_{m}'A'}-2n
	\Phi_{B'A'D(A_{1}}L^{D}{}_{A_{2}\cdots A_{n})B_{1}'\cdots B_{m}'}
	{}^{B'} \nonumber \\Ê& & -2m\bar{\Psi}_{A'B'C'(B_{1}'}L_{|A_{1}\cdots
	A_{n}|B_{2}'\cdots B_{m}')}{}^{B'C'}+2\Lambda\Bigl(3L_{A_{1}\cdots
	A_{n}B_{1}'\cdots B_{m}'A'} \nonumber \\ & & -m\bigl(L_{A_{1}\cdots
	A_{n}A'(B_{1}'\cdots B_{m}')}+\eps_{A'(B_{1}'}L_{|A_{1}\cdots
	A_{n}|B_{2}'\cdots B_{m}')C'}{}^{C'}\bigr)\Bigr) \nonumber \\Ê& &
	+\nabla_{A'}{}^{A}W_{AA_{1}\cdots A_{n}B_{1}'\cdots B_{m}'}-
	\frac{2n}{n+1}\nabla_{A'(A_{1}}F_{A_{2}\cdots A_{n})B_{1}'\cdots B_{m}'}
	\label{mnwave}
\end{eqnarray}
Note that the version of this equation given in \cite{Illge} contains a few
misprints. Suppose we first try to find a completely symmetric solution of
this equation i.e.,
$$
 L_{A_{1}\cdots A_{n}B_{1}'\cdots B_{m}'A'}=L_{(A_{1}\cdots A_{n})(B_{1}'\cdots
 B_{m}'A')}.
$$
Multiplying (\ref{mnwave}) by $\eps^{A'B_{1}'}$ gives
\begin{eqnarray}
	0 & = & n\Phi_{A'B'D(A_{1}}L^{D}{}_{A_{2}\cdots A_{n})B_{2}'\cdots 
	B_{m}'}{}^{A'B'} \nonumber \\ & & +(m-1)\bar{\Psi}_{A'B'C'(B_{2}'}
	L_{|A_{1}\cdots A_{n}|B_{3}'\cdots B_{m}')}{}^{A'B'C'} \nonumber \\
	& & +\frac{1}{2}\nabla^{AA'}W_{AA_{1}\cdots A_{n}A'B_{2}'\cdots B_{m}'}
	-\frac{n}{n+1}\nabla_{(A_{1}}{}^{A'}F_{A_{2}\cdots A_{n})A'B_{2}'\cdots
	B_{m}'}
	\label{mnconstraint}
\end{eqnarray}
Thus, we obtain not only a wave equation for $L$, but also an {\em algebraic 
constraint} on the potential $L$. Therefore we cannot, in general, find a 
completely symmetric potential for a spinor field with both primed and 
unprimed indices. However, we immediately see some cases where these 
constraints are automatically satisfied e.g., when $n=0$, $m=1$ 
providing $\nabla^{AA'}W_{AA'}=0$. Illge \cite{Illge} proves that in 
this case a symmetric potential exists.

Also, if $m=1$ and $\Phi_{ABA'B'}=0$, then the potential vanishes from 
this constraint equation, and we are left with just an equation for 
the differential gauge $F$. If this equation can be solved we might 
expect to find a potential for $W$. These ideas will be explored in 
detail in Section 6. On the other hand, if $n=0$ and $\Psi_{ABCD}=0$ 
we also see that the above equation is no longer a constraint on the 
potential itself.

So, in particular we see that for $W_{AA_{1}\cdots A_{n}A'}$ the 
possibility of having a potential of type $L_{A_{1}\cdots A_{n}A'B'}=
L_{(A_{1}\cdots A_{n})(A'B')}$ in spacetimes with vanishing Ricci spinor, 
is not ruled out. The possibility of $W_{AB_{1}'\cdots B_{m}'}$ 
having a symmetric potential $L_{B_{1}'\cdots B_{m}'A'}$ in 
conformally flat spacetimes is not ruled out either.

Finally we note that if we do {\em not} require complete symmetry of 
$L$, then no constraints occur, and Illge has proven the 
following generalization of Theorem \ref{Illge1} (see \cite{Illge}):

\begin{theorem}\label{Illge2}
 Let symmetric spinor fields $W_{AA_{1}\cdots A_{n}B_{1}'\cdots B_{m}'}$,
 $F_{A_{2}\cdots A_{n}B_{1}'\cdots B_{m}'}$, a spacelike past-compact
 hypersurface $\Sigma$ of class $C^{\infty}$ and a spinor field
 $$
  \ring{L}_{A_{1}\cdots A_{n}B_{1}'\cdots B_{m}'A'}=\ring{L}_{(A_{1}\cdots
  A_{n})(B_{1}'\cdots B_{m}')A'}
 $$
 defined only on $\Sigma$ be given. Then there exists a neighbourhood of
 $\Sigma$ in which the equations
 \begin{eqnarray}
  W_{AA_{1}\cdots A_{n}B_{1}'\cdots B_{m}'} & = & 2\nabla_{(A}{}^{A'}L_{A_{1}
  \cdots A_{n})B_{1}'\cdots B_{m}'A'}\nonumber \\ F_{A_{2}\cdots A_{n}B_{1}'
  \cdots B_{m}'} & = & \nabla^{A_{1}A'}L_{A_{1}A_{2}\cdots A_{n}B_{1}'\cdots
  B_{m}'A'}\nonumber.
 \end{eqnarray}
 have a unique solution $L_{A_{1}\cdots A_{n}B_{1}'\cdots B_{m}'A'}=
 L_{(A_{1}\cdots A_{n})(B_{1}'\cdots B_{m}')A'}$ that satisfies $L|_{\Sigma}=
 \ring{L}$. 
\end{theorem}
In summary, Illge has shown that any symmetric spinor (in fact the 
symmetry condition is not necessary, although these are usually the 
spinors we are interested in) has a potential (actually two different 
potentials using Theorem \ref{Illge2} and the complex conjugate of 
Theorem \ref{Illge2}); but it is only for symmetric spinors which are
restricted to only one type of index where we can {\em always} obtain a
{\em symmetric} potential.

\section{Simple existence proofs for potentials of various spinors}

\subsection{Introduction}

In this section we will give an existence proof for the Lanczos potential
and its generalization. Even though the results of this section can partly
be seen as special cases of Illge's theorem in \cite{Illge}, they have
certain advantages compared to the results in \cite{Illge}. 

The most important advantage is that the existence proof of this section
is conceptually simpler and more direct than the proof given in \cite{Illge}.
This is partly because it is hard to `separate out' the existence part from
the proof in \cite{Illge}. Also the potential $T_{ABCD}$ for the Lanczos 
potential (whose existence could be deduced from the complex 
conjugate of Theorem \ref{Illge2}) turns up as an essential 
part of the theorem. This raises the question whether this potential 
is important in itself.

The obvious drawback of this existence proof is that it is {\em just} 
an existence proof. It does not give us any uniqueness result 
whatsoever.

\subsection{An existence proof for Lanczos potentials}

Let $W_{ABCD}$ and $F_{BC}$ be arbitrary spinor fields. Our objective 
is to show that locally there exists a symmetric spinor $L_{ABCA'}$ 
such that
\begin{equation}
 W_{ABCD}=2\nabla_{(A}{}^{A'}L_{BCD)A'}\;\;,\;\;F_{BC}=\nabla^{AA'}
 L_{ABCA'} \label{WLgauge}
\end{equation}
These equations can be combined into one:
\begin{equation}
	2\nabla_{A}{}^{A'}L_{BCDA'}=W_{ABCD}-\frac{3}{2}\eps_{A(B}F_{CD)}
	\label{WLwgauge}
\end{equation}
Suppose there exists a spinor $T_{ABCD}=T_{(ABC)D}$ such that
\begin{equation}
 L_{ABCA'}=\nabla_{A'}{}^{D}T_{ABCD}
 \label{Tpot}
\end{equation}
where $L_{ABCA'}$ is a solution of (\ref{WLwgauge}). Note that we do 
not invoke (the complex conjugate of) Theorem \ref{Illge2} to ensure
the existence of such a spinor. At the moment we are merely looking at
necessary conditions for its existence. Equation (\ref{WLwgauge}) then
reads
\begin{equation}
	2\nabla_{A}{}^{A^{\prime}}\nabla_{A^{\prime}}{}^{E}T_{BCDE}=
	2\nabla_{A}{}^{A^{\prime}}L_{BCDA^{\prime}}=W_{ABCD}-\frac{3}{2}
	\eps_{A(B}F_{CD)}.
	\label{WLeq}
\end{equation}
On the other hand,
\begin{eqnarray}
	2\nabla_{A}{}^{A^{\prime}}\nabla_{A^{\prime}}{}^{E}T_{BCDE} & = &
	-\Box T_{BCDA}+2\nabla_{A^{\prime}(A}\nabla^{A^{\prime}}{}_{E)}T_{BCD}
	{}^{E} \nonumber \\ & = & -\Box T_{BCDA}-2(X_{AEB}{}^{G}T_{GCD}{}^{E}+
	X_{AEC}{}^{G}T_{BGD}{}^{E} \nonumber \\Ê& & +X_{AED}{}^{G}T_{BCG}{}^{E}-
	X_{AEG}{}^{E}T_{BCD}{}^{G}) \nonumber \\Ê& = & -\Box T_{BCDA}+6
	\Psi_{A(B}{}^{EG}T_{CD)GE}-6\Lambda (\eps_{A(B}T_{CD)E}{}^{E} \nonumber
	\\Ê& & +T_{A(BCD)}+T_{BCDA})
	\nonumber \label{}
\end{eqnarray}
where we have used the commutators \cite{PR1} and the fact that 
$X_{ABCD}=\Psi_{ABCD}+\Lambda(\eps_{AC}\eps_{BD}+\eps_{AD}\eps_{BC})$.
Combining the last equation with (\ref{WLeq}) yields the following wave
equation for $T_{BCDA}$
\begin{eqnarray}
	0 & = & \Box T_{BCDA}-6\Psi_{A(B}{}^{EG}T_{CD)GE}+6\Lambda 
	(\eps_{A(B}T_{CD)E}{}^{E}+T_{A(BCD)}+T_{BCDA})\nonumber \\Ê& &
	-\frac{3}{2}\eps_{A(B}F_{CD)}+W_{ABCD}.
	\label{Twave}
\end{eqnarray}
That this equation is satisfied is a necessary condition for the 
existence of a Lanczos potential of the above type. We will now show
that this equation can also be used to prove the existence of a Lanczos
potential of said type.

\begin{theorem}
 Given symmetric spinors $W_{ABCD}$ and $F_{CD}$ there exists a 
 spinor $L_{ABCA^{\prime}}=L_{(ABC)A^{\prime}}$ satisfying equations 
 (\ref{WLgauge}) locally. Also $L_{ABCA^{\prime}}$ satisfies equation 
 (\ref{Tpot}) for some spinor $T_{ABCD}=T_{(ABC)D}$.
\end{theorem}

\begin{proof}
 Let $p\in M$ be an arbitrary point. From a theorem in 
 \cite{Friedlander} there exists a causal neighbourhood $U$ of $p$. 
 Now, consider the wave equation (\ref{Twave}). This is a linear,
 diagonal second order hyperbolic system for $T_{ABCD}$. Hence, from
 the theory for hyperbolic equations (see e.g., \cite{Wald} or 
 \cite{Friedlander}) we know that it has a solution $T_{ABCD}$ 
 throughout $U$. Put $L_{ABCA^{\prime}}=\nabla_{A^{\prime}}{}^{D}
 T_{ABCD}$; then, because $T_{ABCD}$ is a solution of (\ref{Twave}) it
 follows that
 $$
  2\nabla_{A}{}^{A^{\prime}}L_{BCDA^{\prime}}=2\nabla_{A}{}^{A^{\prime}}
  \nabla_{A^{\prime}}{}^{E}T_{BCDE}=W_{ABCD}-\frac{3}{2}\eps_{A(B}F_{CD)}
 $$
 and as an easy consequence, equation (\ref{WLgauge}) is satisfied. 
\end{proof}

\subsection{Symmetric potentials for $(n+1,0)-$spinors and asymmetric 
potentials for $(n+1,m)-$spinors}

Even though the most studied case is when $W_{ABCD}$ is the Weyl curvature
spinor, there is nothing special about spinors with four indices. Thus, we
immediately obtain the following generalization:

\begin{theorem}
 Given symmetric spinors $W_{AA_{1}\cdots A_{n}}$ and $F_{A_{2}\cdots 
 A_{n}}$ there exists a spinor $L_{A_{1}\cdots A_{n}A^{\prime}}=L_{(A_{1}
 \cdots A_{n})A^{\prime}}$
 satisfying the equations
 \begin{eqnarray}
  W_{AA_{1}\cdots A_{n}} & = & 2\nabla_{(A}{}^{A'}L_{A_{1}\cdots A_{n})A'}
  \nonumber \\ F_{A_{2}\cdots A_{n}} & = & \nabla^{A_{1}A'}L_{A_{1}A_{2}
  \cdots A_{n}A'}\nonumber
 \end{eqnarray}
 locally. Also $L_{A_{1}\cdots A_{n}A^{\prime}}$ satisfies the equation
 $$
  L_{A_{1}\cdots A_{n}A^{\prime}}=\nabla_{A^{\prime}}{}^{B}T_{A_{1}\cdots 
  A_{n}B}
 $$
 for some spinor $T_{A_{1}\cdots A_{n}B}=T_{(A_{1}\cdots A_{n})B}$.
\end{theorem}

\begin{proof}
 It is simply a matter of going through the same calculations as in the
 previous section to arrive at a similar wave equation for $T_{A_{1}\cdots 
 A_{n}B}$ as equation (\ref{Twave}). By the theory for hyperbolic equations,
 this equation will also have a solution locally. Proceed as in the proof 
 of the previous theorem.
\end{proof}

Another possible generalization of the above theorem would be to allow
the spinor $W$ to have primed indices also, and to look for a potential 
having one extra primed index (of course we could reverse the role of
primed and unprimed indices in this argument). Unfortunately it turns out
that if we write down the equation corresponding to equation (\ref{Twave})
it will not necessarily be a linear, diagonal second order hyperbolic
system, if we require our potential to be completely symmetric (see
equations (\ref{mnwave}) and (\ref{mnconstraint})). However, if we remove
the requirement of symmetry over the primed indices, an analogous theorem
can easily be proved in exactly the same way as for the previous theorems.
Thus, to be precise:

\begin{theorem}\label{existprime}
 Given symmetric spinors $W_{AA_{1}\cdots A_{n}B_{1}'\cdots B_{m}'}$ and 
 $F_{A_{2}\cdots A_{n}B_{1}'\cdots B_{m}'}$ there exists a spinor
 $L_{A_{1}\cdots A_{n}B_{1}\cdots B_{m}'B'}=L_{(A_{1}\cdots A_{n})
 (B_{1}\cdots B_{m}')B'}$ satisfying the equations
 \begin{eqnarray}
  W_{AA_{1}\cdots A_{n}B_{1}'\cdots B_{m}'} & = & 2\nabla_{(A}{}^{B'}
  L_{A_{1}\cdots A_{n})B_{1}'\cdots B_{m}'B'}
  \nonumber \\ F_{A_{2}\cdots A_{n}B_{1}'\cdots B_{m}'} & = & 
  \nabla^{A_{1}B'}L_{A_{1} \cdots A_{n}B_{1}'\cdots B_{m}'B'}\nonumber
 \end{eqnarray} 
 locally. Also $L_{A_{1}\cdots A_{n}B_{1}'\cdots B_{m}'B'}$ satisfies the
 equation
 $$
  L_{A_{1}\cdots A_{n}B_{1}'\cdots B_{m}'B'}=\nabla_{B'}{}^{A}T_{A_{1}\cdots 
  A_{n}AB_{1}'\cdots B_{m}'}
 $$
 for some spinor $T_{A_{1}\cdots A_{n}AB_{1}'\cdots B_{m}'}=
 T_{(A_{1}\cdots A_{n})A(B_{1}'\cdots B_{m}')}$.
\end{theorem} 

Note that (the complex conjugate of) this last theorem actually ensures that
for any symmetric spinor $L_{ABCA'}$ there exists a spinor $T_{ABCD}=
T_{(ABC)D}$ such that $L_{ABCA'}=\nabla_{A'}{}^{D}T_{ABCD}$.

We remark once again that the above results only guarantee local 
existence of the Lanczos potential in general. There is however an 
important class of spacetimes for which we can guarantee {\em global} 
existence of the Lanczos potential. If we assume that $M$ has a global 
spinor structure and is globally hyperbolic i.e., contains a Cauchy surface,
then equation (\ref{Twave}) has a global solution $T_{ABCD}$ and if we
put $L_{ABCA^{\prime}}=\nabla_{A^{\prime}}{}^{D}T_{ABCD}$ then 
$L_{ABCA^{\prime}}$ will be globally defined, and will of course still 
be a Lanczos potential. Thus, in globally hyperbolic spacetimes with a 
global spinor structure, the above results guarantee the existence of a
global Lanczos potential.

\subsection{The tensor version}

A spinor with the symmetries $T_{ABCD}=T_{(ABC)D}$ can of course be decomposed 
into $U_{ABCD}=T_{(ABCD)}$ and $V_{AB}=T_{ABC}{}^{C}$. The wave equation
(\ref{Twave}) then splits into
\begin{eqnarray}
	0 & = & \Box U_{ABCD}-6\Psi_{(AB}{}^{EG}U_{CD)GE}+3\Psi_{(ABC}{}^{G}
	V_{D)G}+12\Lambda U_{ABCD}+W_{ABCD} \nonumber \\Ê0 & = & \Box V_{BC}
	-4\Psi^{DEF}{}_{(B}U_{C)DEF}+\Psi_{BC}{}^{DE}V_{DE}-4\Lambda V_{BC}+
	2F_{BC}
	\label{Swavesplit}
\end{eqnarray}
Now, $U_{ABCD}$ corresponds to a tensor $U_{abcd}$ having Weyl symmetry,
and $V_{AB}$ corresponds to a 2-form $V_{ab}$. As before $L_{abc}$ is 
the tensor corresponding to $L_{ABCA'}$. The differential gauge
$F_{ab}$ is defined by $F_{ab}=L_{ab}{}^{c}{}_{;c}$. In this way all the
above definitions carry over to four dimensional spaces of arbitrary 
signature. The above proof can also be directly translated into 
tensors e.g., the wave equations (\ref{Swavesplit}) becomes
\begin{eqnarray}
 \nabla^{2} U_{abcd} & - & \frac{1}{2}(C_{ab}{}^{ef}U_{cdef}+C_{cd}{}^{ef}
 U_{abef})-2C_{c}{}^{e}{}_{[a}{}^{f}U_{b]fde}+2C_{d}{}^{e}{}_{[a}{}^{f}
 U_{b]fce} \nonumber \\ & - & \frac{3}{2}(C_{ab[c}{}^{e}V_{d]e}+C_{cd[a}
 {}^{e}V_{b]e})+\frac{R}{2}U_{abcd}+W_{abcd}=0 \nonumber \\Ê\nabla^{2} V_{bc}
 & + & 4C^{def}{}_{[b}U_{c]def}+\frac{1}{2}C_{bc}{}^{de}V_{de}-\frac{R}{6}
 V_{bc}+2F_{bc}=0 \label{wavesyst}
\end{eqnarray}
Note that here $\nabla^{2}=\nabla^{a}\nabla_{a}$ is not necessarily a wave
operator since $M$ is of arbitrary signature. Now, if $M$ is (real) analytic,
and both $W_{abcd}$ and $F_{bc}$ are (real) analytic then, by the 
Cauchy-Kovalevskaya theorem, this system of equations always has a 
local solution and by translating equation (\ref{Tpot}) into tensors we
can construct a Lanczos potential of $W_{abcd}$ in the differential 
gauge $F_{bc}$ from the solution of (\ref{wavesyst}). We obtain that
\begin{equation}
 L_{abc}=-U_{abc}{}^{d}{}_{;d}-\frac{1}{2}(V_{ab;c}-V_{c[a;b]})-\frac{1}{2}
 g_{c[a}V_{b]}{}^{d}{}_{;d}
 \label{Ltensor}
\end{equation}
is a Lanczos potential of $W_{abcd}$ in the differential gauge $F_{bc}$.

Hence, we have shown that Lanczos potentials exist in 4-dimensional
analytic spacetimes of arbitrary signature in agreement with the
result of Bampi and Caviglia. We remark that this technique seems 
incapable of generalization to spaces of higher dimensions than four.
The reason for this is that when we plug the $n$-dimensional version of 
(\ref{Ltensor}) into the $n$-dimensional version of the Weyl-Lanczos 
equation the resulting equation will not be a wave equation since 
other terms involving second derivatives of $U_{abcd}$ and $V_{bc}$ 
will fail to cancel\footnote{In four dimensions these terms cancel by 
virtue of a special case of Lovelock's identity \cite{Lovelock} in a 
manner analogous to the Lanczos wave equation}. Therefore existence 
of solutions to these equations is not guaranteed. In fact, we have 
strong evidence that Lanczos potentials do not exist, in general, in 
dimensions greater than four \cite{EH2}.

\section{The gauge freedom in the Lanczos potential}

\subsection{General gauge transformations}

The theorems of Section 3 now enable us to characterize the remaining gauge
freedom in the Lanczos potential when the differential gauge is specified.
Let $W_{ABCD}$ and $F_{BC}$ be given symmetric spinors. Let $L_{ABCA'}$ and
$\tilde{L}_{ABCA'}$ be two Lanczos potentials of $W_{ABCD}$ in the same
differential gauge $F_{BC}$ i.e.,
\begin{eqnarray}
	W_{ABCD} & = & 2\nabla_{(A}{}^{A'}L_{BCD)A'}=2\nabla_{(A}{}^{A'}
	\tilde{L}_{BCD)A'} \nonumber \\ F_{BC} & = & \nabla^{AA'}L_{ABCA'}
	=\nabla^{AA'}\tilde{L}_{ABCA'}
	\label{}
\end{eqnarray}
Put $M_{ABCA'}=\tilde{L}_{ABCA'}-L_{ABCA'}$ so that
$$
 2\nabla_{(A}{}^{A'}M_{BCD)A'}=0\;\;,\;\;\nabla^{AA'}M_{ABCA'}=0
$$
which is equivalent to
\begin{equation}
	\nabla_{A}{}^{A'}M_{BCDA'}=0.
	\label{gaugespinor}
\end{equation}
This equation has a formal resemblance to the equation for a 
spin-2-field
$$
 \nabla_{A'}{}^{A}W_{ABCD}=0
$$
which has been studied by Bell and Szekeres \cite{BS}, who found that
in vacuum spacetimes of `sufficient generality' (see \cite{BS}), the only
solutions to this equation are
$$
 W_{ABCD}=c\Psi_{ABCD}
$$
where $c$ is a complex constant. Therefore we might expect 
(\ref{gaugespinor}) to have very few solutions. However, we will see 
that this is {\em not} the case. One reason for this is that by 
taking another derivative and using the commutators, we do not obtain 
any additional algebraic conditions (so called Buchdahl conditions) 
on $M_{ABCA'}$, unlike the very strong condition on $W_{ABCD}$.

Now, according to the complex conjugate of Theorem \ref{existprime} there 
exists a spinor $T_{ABCD}=T_{(ABC)D}$ such that $M_{ABCA'}=\nabla_{A'}
{}^{D}T_{ABCD}$. The same calculations as in the previous section 
tells us that $T_{ABCD}$ must satisfy the following wave equation:
\begin{equation}
	0 =\Box T_{BCDA}-6\Psi_{A(B}{}^{EG}T_{CD)GE}+6\Lambda(\eps_{A(B}
	T_{CD)E}{}^{E}+T_{A(BCD)}+T_{BCDA})
	\label{Twave0}
\end{equation}
or equivalently
\begin{eqnarray}
	0 & = & \Box U_{ABCD}-6\Psi_{(AB}{}^{EG}U_{CD)GE}+3\Psi_{(ABC}{}^{G}
	V_{D)G}+12\Lambda U_{ABCD} \nonumber \\Ê0 & = & \Box V_{BC}
	-4\Psi^{DEF}{}_{(B}U_{C)DEF}+\Psi_{BC}{}^{DE}V_{DE}-4\Lambda V_{BC}
	\label{Twavesplit}
\end{eqnarray}
where $U_{ABCD}=T_{(ABCD)}$ and $V_{BC}=T_{BCD}{}^{D}$. Since these 
equations are coupled we see that in general we need both $U_{ABCD}$ and 
$V_{BC}$ non-zero to get a proper gauge transformation. An important 
exception is when $M$ is conformally flat $(i.e., \Psi_{ABCD}=0)$ where
the equations decouple, and so we could obtain gauge transformations
where e.g., one of $U_{ABCD}$ and $V_{BC}$ is zero, but not the other.
See however Section 4.2.

Thus, we have shown that if $M_{ABCA'}$ constitutes a gauge 
transformation of a Lanczos potential $L_{ABCA'}$ that does not 
change the differential gauge i.e., such that
$$
 \tilde{L}_{ABCA'}=L_{ABCA'}+M_{ABCA'}
$$
is still a Lanczos potential of $W_{ABCD}$ in the differential gauge 
$F_{BC}$ then we can write $M_{ABCA'}=\nabla_{A'}{}^{D}T_{ABCD}$ where
$T_{ABCD}$ is a solution of (\ref{Twave0}). Conversely, suppose $T_{ABCD}$
is a solution of (\ref{Twave0}) and put $M_{ABCA'}=\nabla_{A'}{}^{D}
T_{ABCD}$. Then equation (\ref{Twave0}) can be rewritten as
$$
 0=2\nabla_{A}{}^{A^{\prime}}\nabla_{A^{\prime}}{}^{E}T_{BCDE},
$$
from which
$$
 2\nabla_{A}{}^{A'}M_{BCDA'}=0.
$$
Decomposing into symmetric- and trace parts gives us
$$
 2\nabla_{(A}{}^{A'}M_{BCD)A'}=0\;\;,\;\;\nabla^{AA'}M_{ABCA'}=0
$$
so that $\tilde{L}_{ABCA'}=L_{ABCA'}+M_{ABCA'}$ is also a Lanczos 
potential of $W_{ABCD}$ in the differential gauge $F_{BC}$.

Thus, we have completely characterized the gauge transformations of 
the Lanczos potential that leaves the differential gauge intact. We 
summarize our findings in the following theorem:

\begin{theorem}
 Let $W_{ABCD}$ and $F_{BC}$ be given symmetric spinors. Let $L_{ABCA'}$
 be a Lanczos potential of $W_{ABCD}$ be a Lanczos potential of 
 $W_{ABCD}$ in the differential gauge $F_{BC}$. Then the symmetric 
 spinor $\tilde{L}_{ABCA'}$ is also a Lanczos potential of $W_{ABCD}$ in
 the differential gauge $F_{BC}$ if and only if $\tilde{L}_{ABCA'}=
 L_{ABCA'}+M_{ABCA'}$ where
 $$
  M_{ABCA'}=\nabla_{A'}{}^{D}T_{ABCD}
 $$
 and $T_{ABCD}=T_{(ABC)D}$ is a solution of (\ref{Twave0}).
\end{theorem}

For completeness we also give the tensor version of this result. The 
translation itself is tedious but straightforward.

\begin{theorem}
 Let $W_{abcd}$ be a tensor having all the algebraic symmetries of the 
 Weyl tensor and let $F_{bc}$ be an arbitrary 2-form i.e., 
 $F_{bc}=F_{[bc]}$. Let $L_{abc}$ be a Lanczos potential of $W_{abcd}$ 
 in the differential gauge $F_{bc}$ i.e., $L_{ab}{}^{c}{}_{;c}=F_{bc}$
 Let $\tilde{L}_{abc}$ be a spinor with the same algebraic symmetries 
 of $L_{abc}$. Then $\tilde{L}_{abc}$ is also a Lanczos potential of 
 $W_{abcd}$ in the differential gauge $F_{bc}$ if and only if
 $\tilde{L}_{abc}=L_{abc}+M_{abc}$ where
 $$
  M_{abc}=-U_{abc}{}^{d}{}_{;d}-\frac{1}{2}(V_{ab;c}-
  V_{c[a;b]})-\frac{1}{2}g_{c[a}V_{b]}{}^{d}{}_{;d}
 $$
 and where $U_{abcd}$ have all the algebraic symmetries of the Weyl 
 tensor, $V_{bc}$ is a 2-form and in addition $U_{abcd}$ and $V_{bc}$ 
 satisfies
 \begin{eqnarray}
 \Box U_{abcd} & - & \frac{1}{2}(C_{ab}{}^{ef}U_{cdef}+C_{cd}{}^{ef}
 U_{abef})-2C_{c}{}^{e}{}_{[a}{}^{f}U_{b]fde}+2C_{d}{}^{e}{}_{[a}{}^{f}
 U_{b]fce} \nonumber \\ & - & \frac{3}{2}(C_{ab[c}{}^{e}V_{d]e}+C_{cd[a}
 {}^{e}V_{b]e})+\frac{R}{2}U_{abcd}=0 \nonumber \\Ê\Box V_{bc} & + &
 4C^{def}{}_{[b}U_{c]def}+\frac{1}{2}C_{bc}{}^{de}V_{de}-\frac{R}{6}
 V_{bc}=0 \label{wavesyst2}
\end{eqnarray}
\end{theorem}

\subsection{Gauge transformations with $U_{ABCD}=0$}

In this section we will consider gauge transformations for which 
$U_{ABCD}=0$ i.e., gauge transformations of the form
$$
 \tilde{L}_{ABCA'}=L_{ABCA'}-\frac{3}{4}\nabla_{A'(A}V_{BC)}.
$$
As we saw earlier the chances of finding such gauge transformations, if 
we want to preserve the differential gauge, are rather slim, except in 
the case when $M$ is conformally flat. However, if we allow gauge
transformations that {\em change} the differential gauge the chances are
much better. First note that allowing the differential gauge to change is
the same as solving only the first of equations (\ref{Twavesplit}). When
$U_{ABCD}=0$ this equation becomes
$$
 \Psi_{(ABC}{}^{G}V_{D)G}=0.
$$
Introducing components in the usual way gives us the following system
of equations
\begin{eqnarray}
 0 & = & \Psi_{1}V_{0}-\Psi_{0}V_{1} \nonumber \\ 0 & = & 3\Psi_{2}V_{0}
 -2\Psi_{1}V_{1}-\Psi_{0}V_{2} \nonumber \\ 0 & = & \Psi_{3}V_{0}-
 \Psi_{1}V_{2} \nonumber \\ 0 & = & \Psi_{4}V_{0}+2\Psi_{3}V_{1}-3\Psi_{2}
 V_{2} \nonumber \\Ê0 & = & \Psi_{4}V_{1}-\Psi_{3}V_{2}
 \label{gaugesyst}
\end{eqnarray}
This system can easily be solved for the different Petrov types of 
the Weyl spinor $\Psi_{ABCD}$, using the principal spinors as dyad 
spinors.

The result is that gauge transformations that are allowed to 
change the differential gauge and have $U_{ABCD}=0$ exist, if and only 
if $\Psi_{ABCD}$ is type D, N or 0. In type D we have $V_{AB}=-2V_{1}
o_{(A}\iota_{B)}$ where $o_{A}$ and $\iota_{A}$ are principal spinors 
of $\Psi_{ABCD}$, in type N $V_{AB}=V_{2}o_{A}o_{B}$ where $o_{A}$ is 
the principal spinor of $\Psi_{ABCD}$ and in type 0, $V_{AB}$ is 
arbitrary. We remark that gauge transformations of this type have 
earlier been investigated by Torres del Castillo \cite{TdC1}, 
\cite{TdC2} in the type D and type 0 case. 

It is important to note that in the whole discussion above, $W_{ABCD}$
is arbitrary and therefore, in particular, the results do {\em not} 
depend on the Petrov type of $W_{ABCD}$. They only depend on the 
Petrov type of the Weyl spinor $\Psi_{ABCD}$.

\section{Potentials for symmetric (3,1)-spinors in Einstein spacetimes}

\subsection{Introduction}

In some special cases \cite{lic}, \cite{AE}, \cite{Bergqvist}, \cite{TdC1}
there has been found a completely symmetric spinor $H_{ABA'B'}$ such that
the spinor
$$
 L_{ABCA'}=\nabla_{(A}{}^{B'}H_{BC)A'B'}
$$
is a Lanczos potential of the Weyl spinor. In this section we will prove
that such a spinor $H_{ABA'B'}$ exists in all Einstein spacetimes i.e.,
spacetimes such that the Ricci spinor $\Phi_{ABA'B'}=0$. In fact, we will
prove that in such spacetimes {\em any} symmetric spinor $L_{ABCA'}$ can
be written as
$$
 L_{ABCA'}=\nabla_{(A}{}^{B'}H_{BC)A'B'}
$$
for some spinor $H_{ABA'B'}=H_{(AB)(A'B')}$ and that for each choice 
of $L_{ABCA'}$ there exists many such spinors $H_{ABA'B'}$. We emphasize 
that this result does {\em not} follow from Theorem \ref{Illge2} or 
Theorem \ref{existprime} because here we are requiring {\em complete}
symmetry of $H_{ABA'B'}$.

\subsection{A preliminary result}

First we need a preliminary lemma, which is of interest in its own 
right.

\begin{lemma}\label{vectorlemma}
 For any symmetric spinor field $\varphi_{AB}$, timelike or spacelike
 vector field $n^{AA'}$ and complex function $f$ there exists a unique
 complex vector field $\zeta^{AA'}$ such that $\varphi_{BC}=n_{(B}{}^{A'}
 \zeta_{C)A'}$ and in addition $n^{AA'}\zeta_{AA'}=f$. 
\end{lemma}

\begin{proof}
 By rescaling, it suffices to assume that $n^{AA'}$ is a unit timelike- 
 or spacelike vector. We start by proving uniqueness; suppose that
 $\varphi_{BC}=n_{(B}{}^{A'}\zeta_{C)A'}$ and $n^{AA'}\zeta_{AA'}=f$. For
 the case when $n^{AA'}$ is timelike we obtain, 
 \begin{eqnarray}
 	2\varphi_{A}{}^{B}n_{BA'}+fn_{AA'} & = & n_{A}{}^{B'}\zeta^{B}{}_{B'}
 	n_{BA'}+n^{BB'}\zeta_{AB'}n_{BA'}+fn_{AA'} \nonumber \\ & = &
 	\frac{1}{2}\zeta_{AA'}+n_{AA'}n_{B}{}^{B'}\zeta^{B}{}_{B'}+\eps_{BA}
 	n_{CA'}n^{CB'}\zeta_{B'}{}^{B} \nonumber \\ & & +
 	fn_{AA'} \nonumber \\ & = & \zeta_{AA'}.
 	\label{}
 \end{eqnarray}
 where we have used that $n_{B'}{}^{C}n_{CA'}=\frac{1}{2}\eps_{A'B'}$.
 In the spacelike case, the same calculations give
 $$
  2\varphi_{A}{}^{B}n_{BA'}-fn_{AA'}=\zeta_{AA'}.
 $$
 This proves the uniqueness part so now we need only verify that the 
 above candidate for $\zeta_{AA'}$ actually satisfies the conclusion 
 of the lemma. As before we start with the timelike case:
 $$
  n^{AA'}\bigl(2\varphi_{A}{}^{B}n_{BA'}+fn_{AA'}\bigr)=2\varphi_{A}{}^{B}
  \cdot \frac{1}{2}\eps_{B}{}^{A}+f=f
 $$
 and
 $$
  n_{(B}{}^{A'}\zeta_{C)A'}=2n_{(B}{}^{A'}\varphi_{C)}{}^{D}n_{DA'}+f
  n_{(B}{}^{A'}n_{C)A'}=\eps_{D(B}\varphi_{C)}{}^{D}=\varphi_{BC}
 $$
 This proves the lemma in the timelike case. The spacelike case is 
 proved in exactly the same way.
\end{proof}

\subsection{Construction of the spinor potential}

Let $M$ be an Einstein spacetime i.e., $\Phi_{ABA'B'}=0$ and let
$L_{ABCA^{\prime}}$ be a symmetric spinor field on $M$. Our objective is to
show that locally there exists a spinor field $H_{ABA^{\prime}B^{\prime}}=
H_{(AB)(A^{\prime}B^{\prime})}$ such that
\begin{equation}
 L_{ABCA^{\prime}}=\nabla_{(A}{}^{B^{\prime}}H_{BC)A^{\prime}B^{\prime}}.
 \label{Heqn}
\end{equation}
We also wish to examine the gauge freedom in the potential
$H_{ABA^{\prime}B^{\prime}}$. 

Our strategy for proving the existence of $H_{ABA^{\prime}B^{\prime}}$ 
will be to start by deriving a wave equation for $H_{ABA^{\prime}B^{\prime}}$,
along with some constraint equations. Then we use the same theorem 
from \cite{Friedlander} as in Section 3 to show that these equations have
a solution; finally we prove that this solution also solves equation
(\ref{Heqn}).

We begin by assuming that $H_{ABA^{\prime}B^{\prime}}=
H_{(AB)(A^{\prime}B^{\prime})}$ satisfies
\begin{equation}
 	\nabla_{(A}{}^{B^{\prime}}H_{BC)A^{\prime}B^{\prime}}=L_{ABCA^{\prime}}
 	\quad , \quad \nabla^{AA^{\prime}}H_{ABA^{\prime}B^{\prime}}=
 	\zeta_{BB^{\prime}}
 	\label{Heqn2}
\end{equation}
where $\zeta_{BB^{\prime}}$ is a given spinor field (complex 1-form). Note
that
$$
 \nabla_{A}{}^{B^{\prime}}H_{BCA^{\prime}B^{\prime}}=\nabla_{(A}
 {}^{B^{\prime}}H_{BC)A^{\prime}B^{\prime}}-\frac{2}{3}\eps_{A(B}
 \nabla^{DD^{\prime}}H_{C)DA^{\prime}D^{\prime}}
$$ 
so (\ref{Heqn2}) is equivalent to
\begin{equation}
	\nabla_{A}{}^{B^{\prime}}H_{BCA^{\prime}B^{\prime}}=L_{ABCA^{\prime}}-
	\frac{2}{3}\eps_{A(B}\zeta_{C)A^{\prime}}
	\label{Heqn3}
\end{equation}
Now, let $\Sigma$ be a $C^{\infty}$ spacelike past-compact hypersurface
with future-directed unit normal $n^{a}=n^{AA^{\prime}}$. Let $\nabla_{n}
=n^{a}\nabla_{a}=n^{AA^{\prime}}\nabla_{AA^{\prime}}$ be the normal 
derivative with respect to $\Sigma$ and let $\tilde{\nabla}_{AA^{\prime}}
=\nabla_{AA^{\prime}}-n_{AA^{\prime}}\nabla_{n}$ so that
$\tilde{\nabla}_{AA^{\prime}}$ is the part of $\nabla_{AA^{\prime}}$ that
acts tangentially to $\Sigma$.

Put $\ring{H}_{ABA^{\prime}B^{\prime}}=
H_{ABA^{\prime}B^{\prime}}|_{\Sigma}$. Since (\ref{Heqn3}) must be
satisfied also on $\Sigma$ we obtain
$$
 n_{A}{}^{B^{\prime}}\nabla_{n}H_{BCA^{\prime}B^{\prime}}|_{\Sigma}=
 -\bigl(\tilde{\nabla}_{A}{}^{B^{\prime}}H_{BCA^{\prime}B^{\prime}}
 -L_{ABCA^{\prime}}+\frac{2}{3}\eps_{A(B}\zeta_{C)A^{\prime}}\bigr)
 |_{\Sigma}.
$$
As before we have that $n^{AC^{\prime}}n_{A}{}^{B^{\prime}}=\frac{1}{2}
\eps^{B^{\prime}C^{\prime}}$. Thus, multiplying the previous equation by
$n^{AC^{\prime}}$ gives us an explicit expression for the normal derivative
of $H_{ABA^{\prime}B^{\prime}}$.
\begin{equation}
	\nabla_{n}H_{BCA^{\prime}}{}^{C^{\prime}}|_{\Sigma}=2n^{AC^{\prime}}
	\bigl(\tilde{\nabla}_{A}{}^{B^{\prime}}H_{BCA^{\prime}B^{\prime}}-
	L_{ABCA^{\prime}}+\frac{2}{3}\eps_{A(B}\zeta_{C)A^{\prime}}\bigr)
	|_{\Sigma}
	\label{}
\end{equation}
Note that since $\tilde{\nabla}_{AA^{\prime}}$ only consists of 
derivatives in directions tangential to $\Sigma$ we can replace $H$ with
$\ring{H}$ in the RHS. If we lower the index $C^{\prime}$ then the LHS is
symmetric over $(A^{\prime}C^{\prime})$. Hence, the above equation is
equivalent to the following initial value constraints:
\begin{eqnarray}
	\nabla_{n}H_{BCA^{\prime}C^{\prime}}|_{\Sigma} & = & 2n^{A}
	{}_{(C^{\prime}}\bigl(\tilde{\nabla}_{|A}{}^{B^{\prime}}
	\ring{H}_{BC|A^{\prime})B^{\prime}}-L_{|ABC|A^{\prime})}+
	\frac{2}{3}\eps_{|A(B}\ring{\zeta}_{C)|A^{\prime})}\bigr)|_{\Sigma}
	\nonumber \\Ê0 & = & n^{AA^{\prime}}\bigl(\tilde{\nabla}_{A}
	{}^{B^{\prime}}\ring{H}_{BCA^{\prime}B^{\prime}}-L_{ABCA^{\prime}}+
	\frac{2}{3}\eps_{A(B}\ring{\zeta}_{C)A^{\prime}}\bigr)|_{\Sigma}.
	\label{constraint}
\end{eqnarray}
where we have put $\ring{\zeta}_{AA^{\prime}}=\zeta_{AA^{\prime}}
|_{\Sigma}$.

Next we differentiate the LHS of (\ref{Heqn3}):
\begin{eqnarray}
	\nabla^{A}{}_{C^{\prime}}\nabla_{A}{}^{B^{\prime}}
	H_{BCA^{\prime}B^{\prime}} & = & \eps^{B^{\prime}D^{\prime}}\nabla^{A}
	{}_{(C^{\prime}}\nabla_{D^{\prime})A}H_{BCA^{\prime}B^{\prime}}
	+\frac{1}{2}\nabla^{A}{}_{E^{\prime}}\nabla_{A}{}^{E^{\prime}}
	H_{BCA^{\prime}C^{\prime}} \nonumber \\ & = & -\frac{1}{2}\Box
	H_{BCA^{\prime}C^{\prime}}+\bar{\Psi}_{B^{\prime}E^{\prime}A^{\prime}
	C^{\prime}}H_{BC}{}^{E^{\prime}B^{\prime}}\nonumber \\ & & -4\Lambda
	H_{BCA'C'}
	\label{}
\end{eqnarray}
where we have used that $\Phi_{ABA'B'}=0$ along with the symmetry of 
$H_{ABA'B'}$. Thus, $H_{ABA^{\prime}B^{\prime}}$ satisfies the following
wave equation:
\begin{eqnarray}
	\Box H_{BCA^{\prime}C^{\prime}} & - & 2\bar{\Psi}_{B^{\prime}E^{\prime}
	A^{\prime}C^{\prime}}H_{BC}{}^{E^{\prime}B^{\prime}}+8\Lambda 
	H_{BCA'C'} \nonumber \\Ê& = & -2\nabla^{A}{}_{C^{\prime}}L_{ABCA^{\prime}}
	+\frac{4}{3}\nabla_{C^{\prime}(B}\zeta_{C)A^{\prime}}
	\label{waveeqn1}
\end{eqnarray}
Note that this equation is actually a special case of equation (\ref{mnwave})
of Section 2.
 
Since $H_{BCA^{\prime}C^{\prime}}$ is symmetric over 
$(A^{\prime}C^{\prime})$ it follows that (\ref{waveeqn1}) is 
equivalent to
\begin{eqnarray}
	\Box H_{BCA^{\prime}C^{\prime}} & - & 2
	\bar{\Psi}_{B^{\prime}E^{\prime}A^{\prime}C^{\prime}}
	H_{BC}{}^{E^{\prime}B^{\prime}}+8\Lambda H_{BCA'C'} \nonumber 
	\\ & = & -2\nabla^{A}{}_{(C^{\prime}}L_{|ABC|A^{\prime})}+ 
	\frac{2}{3}\nabla_{C^{\prime}(B}\zeta_{C)A^{\prime}}+\frac{2}{3}
	\nabla_{A^{\prime}(B}\zeta_{C)C^{\prime}} \nonumber \\ 0 & = & 
	-\nabla^{AA^{\prime}}L_{ABCA^{\prime}}+\frac{2}{3}
	\nabla^{A^{\prime}}{}_{(B}\zeta_{C)A^{\prime}}
	\label{waveeqn2}
\end{eqnarray}
The second of these equations is actually equation (\ref{mnconstraint}) 
of Section 2.

After these preliminary considerations we are ready to prove our main 
result.

\begin{theorem}
 Suppose $M$ is an Einstein spacetime $(\Phi_{ABA'B'}=0)$ and that
 $\Sigma\subset M$ is a $C^{\infty}$ spacelike past-compact hypersurface
 with future directed unit normal $n^{AA^{\prime}}$. Let a spinor field 
 $L_{ABCA^{\prime}}=L_{(ABC)A^{\prime}}$ and a complex function $g$ be
 given. Furthermore, let a spinor field $\ring{H}_{ABA^{\prime}B^{\prime}}
 =\ring{H}_{(AB)(A^{\prime}B^{\prime})}$ and a complex function $\ring{f}$,
 both defined only on $\Sigma$ be given. Then there exists a neighbourhood
 $U$ of $\Sigma$ such that there exists a unique spinor field
 $H_{ABA^{\prime}B^{\prime}}=H_{(AB)(A^{\prime}B^{\prime})}$ satisfying the
 equations
 \begin{eqnarray}
 	\nabla_{(A}{}^{B^{\prime}}H_{BC)A^{\prime}B^{\prime}} & = &
 	L_{ABCA^{\prime}} \nonumber \\Ê\nabla^{AA^{\prime}}
 	\nabla^{BB^{\prime}}H_{ABA^{\prime}B^{\prime}} & = & g 
 	\nonumber \\ H_{ABA^{\prime}B^{\prime}}|_{\Sigma} & = &
 	\ring{H}_{ABA^{\prime}B^{\prime}} \nonumber \\Ên^{AA^{\prime}}
 	\nabla^{BB^{\prime}}H_{ABA^{\prime}B^{\prime}}|_{\Sigma} & = &
 	\ring{f}
 	\label{conditions}
 \end{eqnarray}
 on all of $U$.
\end{theorem}

\begin{proof}
 An outline of the existence part of the proof is as follows. We start by
 solving the second of the equations (\ref{constraint}) for
 $\ring{\zeta}_{AA^{\prime}}$ so that $n^{AA'}\ring{\zeta}_{AA'}=\ring{f}$.
 Then we evolve this initial data using the second equation of 
 (\ref{waveeqn2}) in such a way that $\nabla^{AA'}\zeta_{AA'}=g$. Next
 we calculate the normal derivative of $H_{ABA^{\prime}B^{\prime}}$ using
 the first equation of (\ref{constraint}) and use the so obtained Cauchy
 data for $H_{ABA^{\prime}B^{\prime}}$ to solve the first equation of 
 (\ref{waveeqn2}) for $H_{ABA^{\prime}B^{\prime}}$. It can then be 
 verified that this spinor field satisfies all the conditions of the 
 theorem. 
 
 Define the symmetric spinor
 $$
  \ring{\varphi}_{BC}=\frac{3}{2}n^{AA^{\prime}}\bigl(L_{ABCA^{\prime}}
  |_{\Sigma}-\tilde{\nabla}_{A}{}^{B^{\prime}}
  \ring{H}_{BCA^{\prime}B^{\prime}}\bigr)
 $$
 By Lemma \ref{vectorlemma} there exists a unique spinor 
 $\ring{\zeta}_{AA^{\prime}}$ such that
 $$
  n^{AA'}\ring{\zeta}_{AA'}=\ring{f}
 $$
 and such that the second of the equations (\ref{constraint}) is satisfied
 i.e.,
 \begin{equation}
  \ring{\varphi}_{BC}=n_{(B}{}^{A'}\ring{\zeta}_{C)A'}.
  \label{constraint2}
 \end{equation}
 Our next task will be to solve for $\zeta_{AA'}$. We want to find 
 $\zeta_{AA'}$ so that the following three conditions are satisfied
 \begin{eqnarray}
 	\nabla_{(B}{}^{A'}\zeta_{C)A'} & = & \frac{3}{2}\nabla^{AA'}L_{ABCA'}
 	\nonumber \\ \nabla^{AA'}\zeta_{AA'} & = & g \nonumber \\
 	\zeta_{AA'}|_{\Sigma} & = & \ring{\zeta}_{AA'}
 	\label{}
 \end{eqnarray}
 where $\ring{\zeta}_{AA'}$ is the solution of (\ref{constraint2}) 
 obtained above. Let $U$ be a causal neighbourhood \cite{Friedlander} 
 of $\Sigma$. According to Theorem \ref{Illge1} this problem
 has a unique solution $\zeta_{AA'}$ in $U$.
 
 Next, consider the problem
 \begin{eqnarray}
 	\Box H_{BCA^{\prime}C^{\prime}} & - & 2\bar{\Psi}_{B^{\prime}E^{\prime}
	A^{\prime}C^{\prime}}H_{BC}{}^{E^{\prime}B^{\prime}}+8\Lambda 
	H_{BCA'C'} \nonumber \\ & = & -2\nabla^{A}{}_{(C^{\prime}}
	L_{|ABC|A^{\prime})}+\frac{2}{3}\nabla_{C^{\prime}(B}
	\zeta_{C)A^{\prime}}+\frac{2}{3}\nabla_{A^{\prime}(B}\zeta_{C)C^{\prime}}
	\nonumber \\ \nabla_{n}H_{BCA^{\prime}C^{\prime}}|_{\Sigma} & = & 2
	n^{A}{}_{(C'}\bigl(\tilde{\nabla}_{|A}{}^{B^{\prime}}
	\ring{H}_{BC|A^{\prime})B^{\prime}}-L_{|ABC|A^{\prime})} \nonumber \\
	& & +\frac{2}{3}\eps_{|A(B}\ring{\zeta}_{C)|A^{\prime})}\bigr)|_{\Sigma}
	\nonumber \\ H_{BCA'C'}|_{\Sigma} & = & \ring{H}_{BCA'C'}.
 	\label{}
 \end{eqnarray}
 These are the first equation of (\ref{waveeqn2}), the first equation 
 of (\ref{constraint}) and the third condition of (\ref{conditions}). 
 Note that the RHS of all three equations contain only known 
 quantities. Hence this problem is a Cauchy problem for a linear, 
 diagonal, second order hyperbolic system. According to a theorem in 
 \cite{Friedlander} and \cite{Wald} this problem has a unique solution
 $H_{BCA'C'}$ in $U$.
 
 It now remains to prove that the $H_{BCA'C'}$ found above satisfies 
 the conditions
 $$
  \nabla_{(A}{}^{B'}H_{BC)A'B'}=L_{ABCA'}\quad , \quad \nabla^{BB'}
  H_{ABA'B'}=\zeta_{AA'}.
 $$
 In order to do that we define
 $$
  \xi_{ABCA'}=\nabla_{A}{}^{B^{\prime}}H_{BCA^{\prime}B^{\prime}}-
  L_{ABCA^{\prime}}+\frac{2}{3}\eps_{A(B}\zeta_{C)A^{\prime}}.
 $$
 Equation (\ref{constraint}) now implies that $\xi_{ABCA'}|_{\Sigma}=0$.
 Because both $H_{BCA'C'}$ and $\zeta_{AA'}$ are constructed so that 
 equation (\ref{waveeqn2}) are satisfied we have that $\nabla^{A}{}_{C'}
 \xi_{ABCA'}=0$. Because $\tilde{\nabla}_{AA'}$ only consists of 
 derivatives in directions tangential to $\Sigma$, this gives us that
 $$
  n^{A}{}_{C'}\nabla_{n}\xi_{ABCA'}|_{\Sigma}=-\bigl(\tilde{\nabla}^{A}{}_{C'}
  \xi_{ABCA'}\bigr)|_{\Sigma}=-\tilde{\nabla}^{A}{}_{C'}\bigl(
  \xi_{ABCA'}|_{\Sigma}\bigr)=0.
 $$
 Thus,
 $$
  0=n^{DC'}n^{A}{}_{C'}\nabla_{n}\xi_{ABCA'}|_{\Sigma}=\frac{1}{2}
  \eps^{AD}\nabla_{n}\xi_{ABCA'}|_{\Sigma}=-\frac{1}{2}\nabla_{n}
  \xi^{D}{}_{BCA'}
 $$
 Taking another derivative gives us
 \begin{eqnarray}
 	0 & = & \nabla_{D}{}^{C'}\nabla^{A}{}_{C'}\xi_{ABCA'} \nonumber \\
 	& = & -\frac{1}{2}\Box \xi_{DBCA'}+2\Psi_{D(B}{}^{AF}\xi_{|A|C)FA'}-
 	3\Lambda \xi_{DBCA'}-2\Lambda\xi_{(BC)DA'} \nonumber \\ & & -2\Lambda
 	\eps_{D(B}\xi^{A}{}_{C)AA'}
  	\label{}
 \end{eqnarray}
 because we assumed that $M$ is Einstein. Hence $\xi_{ABCA'}$ is a solution of
 the following problem
 \begin{eqnarray}
 	\Box \xi_{DBCA'}-4\Psi_{D(B}{}^{AF}\xi_{|A|C)FA'}+6\Lambda\xi_{DBCA'} & & 
 	\nonumber \\ +4\Lambda\xi_{(BC)DA'}+4\Lambda\eps_{D(B}\xi^{A}{}_{C)AA'}
 	& = & 0 \nonumber \\Ê\xi_{DBCA'}|_{\Sigma} & = & 0 \nonumber \\Ê\nabla_{n}
 	\xi_{DBCA'}|_{\Sigma} & = & 0
 	\label{}
 \end{eqnarray}
 This homogeneous problem has a unique solution in $U$ according to
 \cite{Friedlander}. Therefore we must have
 $$
  \xi_{ABCA'}=0
 $$
 in $U$, which implies that
 $$
  \nabla_{(A}{}^{B'}H_{BC)A'B'}=L_{ABCA'}\quad , \quad \nabla^{BB'}
  H_{ABA'B'}=\zeta_{AA'}.
 $$
 This proves that $H_{BCA'C'}$ satisfies all the conditions 
 (\ref{conditions}), which completes the existence part of the theorem.
 
 {\em Uniqueness:} Remember that $\ring{\zeta}_{AA'}$ was
 uniquely determined by the fourth condition of (\ref{conditions}) and
 the second equation of (\ref{constraint}) and that $\zeta_{AA'}$ was
 uniquely determined by $\ring{\zeta}_{AA'}$, the second condition of
 (\ref{conditions}) and the second equation of (\ref{waveeqn2}). Also 
 recall that this determined the normal derivative of $H_{BCA'C'}$ 
 on $\Sigma$ uniquely and that this normal derivative together with the 
 third condition of (\ref{conditions}) and the first equation of 
 (\ref{waveeqn2}) determined $H_{BCA'C'}$ uniquely. This establishes
 uniqueness.
 
\end{proof}

\subsection{The tensor potential}

It is tedious but straightforward to translate the above result into 
tensors. The condition $\Phi_{ABA'B'}=0$ translates into the vanishing 
of the trace-free Ricci tensor $\tilde{R}_{ab}=R_{ab}-\frac{1}{4}Rg_{ab}$
and as mentioned above, $L_{ABCA'}$ corresponds to a real tensor $L_{abc}$
such that
$$
 L_{abc}=L_{[ab]c}\;\;,\;\;L_{[abc]}=0\;\;,\;\;L_{ab}{}^{b}=0
$$
We also note that a spinor field $H_{ABA'B'}=H_{(AB)(A'B')}$ 
corresponds to a complex, symmetric and trace-free tensor field $H_{ab}$
i.e.,
$$
 H_{ab}=H_{(ab)}\;\;,\;\;H_{a}{}^{a}=0
$$
\begin{theorem}
 Suppose $M$ is an Einstein spacetime $(\tilde{R}_{ab}=0)$ and that
 $\Sigma\subset M$ is a $C^{\infty}$ spacelike hypersurface with future
 directed unit normal $n^{a}$. Let a real tensor field $L_{abc}$ 
 having the above symmetries, and a complex function $g$ be given.
 Furthermore, let a complex function $\ring{f}$ and a complex tensor
 field $\ring{H}_{ab}=\ring{H}_{(ab)}$ such that $\ring{H}_{a}{}^{a}=0$,
 both defined only on $\Sigma$ be given. Then there exists a neighbourhood
 $U$ of $\Sigma$ such that there exists a unique complex tensor field
 $H_{ab}$ satisfying the equations
 \begin{eqnarray}
    H_{ab} & = & H_{(ab)}\;\;,\;\;H_{a}{}^{a}=0 \nonumber \\
 	L_{abc} & = & -\nabla_{[a}H_{b]c}-\nabla_{[a}\bar{H}_{b]c}-i
 	\nabla^{*}_{[a}H_{b]c}+i\nabla^{*}_{[a}\bar{H}_{b]c}+\frac{1}{3}
 	\Bigl(g_{c[a}\nabla^{d}H_{b]d} \nonumber \\ & & +g_{c[a}\nabla^{d}
 	\bar{H}_{b]d}+ig^{*}_{c[a}\nabla^{d}H_{b]d}-ig^{*}_{c[a}\nabla^{d}
 	\bar{H}_{b]d}\Bigr)\nonumber \\Ê\nabla^{a}\nabla^{b}H_{ab} & = & g 
 	\nonumber \\ H_{ab}|_{\Sigma} & = & \ring{H}_{ab} \nonumber \\Ên^{a}
 	\nabla^{b}H_{ab}|_{\Sigma} & = & \ring{f}
 	\label{tensconditions}
 \end{eqnarray}
 on all of $U$.
\end{theorem}
By writing $H_{ab}=H^{1}_{ab}+iH^{2}_{ab}$ where $H^{1}$ and $H^{2}$ 
are real, we can simplify the second of the above conditions somewhat
\begin{equation}
 L_{abc}=-2\nabla_{[a}H^{1}_{b]c}+2\nabla^{*}_{[a}H^{2}_{b]c}
 +\frac{2}{3}\Bigl(g_{c[a}\nabla^{d}H^{1}_{b]d}-g^{*}_{c[a}\nabla^{d}
 H^{2}_{b]d}\Bigr). \label{H1H2}
\end{equation} 
This theorem can be generalized to four dimensional {\em analytic} 
spaces of arbitrary signature, just like the theorem in Section 3.

\section{Comparison with electromagnetic theory}

\subsection{Introduction}

In this section we consider electromagnetic theory in a curved 
spacetime. Most of the above results are applicable here too, and
we will also find that due to the simple index configuration 
of the electromagnetic spinor ($\varphi_{AB}$ as compared to 
$\Psi_{ABCD}$) and also due to Maxwell's equations, certain 
simplifications will occur.

\subsection{The electromagnetic field and its spinor potentials}

First of all we remark that as in the rest of the paper all results 
in this section are local in nature unless comments are made to the
contrary.

Recall that the electromagnetic tensor (Maxwell tensor) is a 2-form 
$F_{ab}=F_{[ab]}$. Maxwell's equations are
$$
 \nabla^{a}F_{ab}=J_{b}\;,\;\nabla_{[a}F_{bc]}=0
$$
where $J_{b}$ is the source current. The second of these equations together
with Poincare's lemma gives us the existence of a (real) 1-form $A_{a}$ such
that
$$
 F_{ab}=\nabla_{[a}A_{b]}.
$$
Now, put $\alpha=\nabla^{a}A_{a}$ (i.e., $\alpha$ is analogous to the 
differential gauge in the above sections). If $\alpha=0$ the electromagnetic 
potential $A_{a}$ is said to be in Lorenz gauge.

To examine the gauge freedom in $A_{a}$, suppose $A_{a}$ and 
$\tilde{A}_{a}$ are two potentials of $F_{ab}$ in the same 
differential gauge and put $B_{a}=\tilde{A}_{a}-A_{a}$. Then
\begin{equation}
 \nabla_{[a}B_{b]}=0\;,\;\nabla^{a}B_{a}=0.
 \label{EMgauge}
\end{equation}
Thus, there exists a (real) scalar field $G$ such that 
$B_{a}=\nabla_{a}G$ and by the second condition then $\Box G=0$.

Conversely, take any scalar field $G$ that satisfies $\Box G=0$ and 
put $B_{a}=\nabla_{a}G$. Then $B_{a}$ satisfies equation 
(\ref{EMgauge}) and therefore $\tilde{A}_{a}=A_{a}+B_{a}$ will be a 
potential of $F_{ab}$ in the same differential gauge as $A_{a}$. 
Hence, we have completely characterized the gauge transformations 
that preserve the differential gauge.

Next we turn to the spinor formulation. As $F_{ab}$ is antisymmetric 
it can be written
$$
 F_{ab}=\varphi_{AB}\eps_{A'B'}+\overline{\varphi}_{A'B'}\eps_{AB}
$$
for some symmetric spinor $\varphi_{AB}$. Maxwell's equations can be 
shown to be
$$
 \nabla_{A'}{}^{B}\varphi_{AB}=J_{AA'}
$$
where $J_{AA'}$ is the hermitian spinor equivalent of the current $J_{a}$. 
If we apply Illge's Theorem 2.2 to $\varphi_{AB}$ we obtain the existence 
of a {\em complex} 1-form $A_{AA'}$ such that
\begin{equation}
 \varphi_{AB}=\nabla_{(A}{}^{A'}A_{B)A'}
 \label{Aphi}
\end{equation}
Putting $A_{AA'}=-\frac{1}{2}A^{1}_{AA'}+\frac{i}{2}A^{2}_{AA'}$ 
where $A^{i}_{AA'}\,,\,i=1,2$ are hermitian this equation becomes (in 
tensors) \cite{Illge}
$$
 F_{ab}=\nabla_{[a}A^{1}_{b]}+{}^{*}\nabla_{[a}A^{2}_{b]}
$$
where ${}^{*}$ denotes the Hodge dual. It is shown in \cite{Illge} 
that solutions of this equation, with $A^{2}_{a}=0$ exist only if 
$\nabla_{[a}F_{bc]}=0$ (which is true if and only if $J_{AA'}$ is 
hermitian) in agreement with Poincare's lemma.

It is interesting to note that the existence of the potential 
$A_{a}$ in electromagnetic theory is usually presented as a 
consequence of the second of Maxwell's equations via Poincare's 
lemma. However we see that the existence of the (complex) potential 
$A_{AA'}$ is independent of Maxwell's equations; it is simply a 
consequence of Theorem 2.2. The role of Maxwell's equations is to 
ensure that this potential is hermitian.

Now, we can of course use the theorems in the earlier sections to 
find potentials of $A_{AA'}$. From Theorem 2.2  (or 3.3) we know that
we can always find an asymmetric potential $H_{A'B'}$ (however,
when $A_{AA'}$ is divergence-free i.e., $\alpha=0$ it is shown in 
\cite{Illge} that a symmetric potential always exists, see also below)
and from the complex conjugate of Theorem 2.2 (or 3.3) we can obtain an
asymmetric potential $T_{AB}$. So we have two potentials for $A_{AA'}$
satisfying
$$
 A_{AA'}=\nabla_{A'}{}^{B}T_{AB}=\nabla_{A}{}^{B'}H_{A'B'}.
$$
It is easily seen that if $A_{AA'}$ is hermitian then if $T_{AB}$ is a 
potential of $A_{AA'}$ then $H_{A'B'}=\overline{T}_{A'B'}$ is also a
potential of $A_{AA'}$.

It is to be noted that if $F_{ab}$ does {\em not} satisfy
Maxwell's equations then we cannot choose the electromagnetic 
potential $A_{AA'}$ hermitian, and there is no simple relation between 
the two potentials $T_{AB}$ and $H_{A'B'}$.

As before we can also obtain a wave equation for $T_{AB}$. Decomposed 
into its symmetric and antisymmetric parts it becomes
\begin{eqnarray}
 0 & = & \Box T_{(AB)}-2\Psi_{AB}{}^{CD}T_{(CD)}+8\Lambda T_{(AB)}+
 2\varphi_{AB} \nonumber \\ 0 & = & \Box T_{A}{}^{A}+2\alpha
\end{eqnarray}
highlighting a formal resemblance between $T_{AB}$ and the Hertz potential
in flat space.

As in Section 4 we can express the gauge freedom of $A_{AA'}$ in 
terms of $T_{AB}$. The result is that $A_{AA'}$ and $\tilde{A}_{AA'}
=A_{AA'}+B_{AA'}$ are two potentials of $\varphi_{AB}$ in the 
differential gauge $\alpha$ if and only if
$$
 B_{AA'}=\nabla_{A'}{}^{B}T_{AB}
$$
where $T_{AB}$ is a solution of
\begin{eqnarray}
 0 & = & \Box T_{(AB)}-2\Psi_{AB}{}^{CD}T_{(CD)}+8\Lambda T_{(AB)} \nonumber
 \\ 0 & = & \Box T_{A}{}^{A} \label{EMgaugeT}
\end{eqnarray}
But we had already expressed the gauge freedom in terms of the 
scalar $G$, so we might wonder what the link between $T_{AB}$ and $G$ 
is. To give a partial answer to this question, let $\Sigma$ be as in 
Section 5 and suppose
$$
 B_{AA'}=\nabla_{A'}{}^{B}T_{AB}=\nabla_{AA'}G.
$$
where $T_{AB}$ satisfies the first of equations (\ref{EMgaugeT}) and 
$G$ is an arbitrary scalar field (so that the gauge transformation 
$B_{AA'}$ is allowed to change the differential gauge). 
It follows that
$$
 0=\nabla_{A'}{}^{B}(T_{AB}+\eps_{AB}G).
$$
By differentiating again we obtain ($S_{AB}=T_{(AB)}$, $T=T_{A}{}^{A}$)
\begin{eqnarray}
 0 & = & \Box S_{AB}-2\Psi_{AB}{}^{CD}S_{CD}+8\Lambda S_{AB} \nonumber
 \\ 0 & = & \Box (T+2G)
\end{eqnarray}
and by evaluating on $\Sigma$ we get
\begin{eqnarray}
 \nabla_{n}S_{AC}|_{\Sigma} & = & \Bigl(-2n_{A'(C}\tilde{\nabla}^{A'B}
 S_{A)B}+n_{A'(C}\tilde{\nabla}^{A'}{}_{A)}T\Bigr)|_{\Sigma} \nonumber
 \\ \nabla_{n}(T+2G) & = & 2n^{AA'}\tilde{\nabla}_{A'}{}^{B}
 S_{AB}|_{\Sigma}
\end{eqnarray}
It easily follows that if $T|_{\Sigma}=-2G|_{\Sigma}$ and if $n^{AA'}
\tilde{\nabla}_{A'}{}^{B} S_{AB}|_{\Sigma}=0$ then $T=-2G$ in a 
neighbourhood of $\Sigma$.

Finally we will look a little closer at the case when $F_{ab}$ is a
2-form that satisfy Maxwell's equations. Poincare's lemma (or
\cite{Illge}) then tells us that there exists a (hermitian) divergence-free
potential $A_{a}=A_{AA'}$. Now, according to Illge \cite{Illge} for any
{\em complex} divergence-free 1-form $A_{AA'}$ there exists a
{\em symmetric} spinor $T_{AB}$ such that $A_{AA'}=\nabla_{A'}{}^{B}
T_{AB}$. Define the 2-form $T_{ab}=T_{AB}\eps_{A'B'}+\overline{T}_{A'B'}
\eps_{AB}$. The tensor equations relating $A_{a}$ and $T_{ab}$ are 
then
$$
 \nabla^{a}T_{ab}=2\,Re(A_{a})\;,\;{}^{*}\nabla^{a}T_{ab}=2\,Im(A_{a})
$$
As $A_{a}$ was chosen hermitian we obtain
$$
 \nabla^{a}T_{ab}=2A_{a}\;,\;{}^{*}\nabla^{a}T_{ab}=0
$$
The second equation of these is equivalent to $\nabla_{[a}T_{bc]}$ 
i.e., $T_{ab}$ is a closed 2-form just like $F_{ab}$ so $T_{ab}$ also has
a hermitian, divergence-free potential and so on. Hence, we get an 
infinite chain of potentials alternating between hermitian, 
divergence-free 1-forms and closed hermitian 2-forms.

\section{Discussion}

The most important motivation for studying the general spinor 
potentials of the earlier sections has been the Lanczos potential of 
the Weyl curvature spinor. The discussion of this section will 
therefore deal mainly with those potentials and their `superpotentials'
$H_{ABA'B'}$ and $T_{ABCD}$.

Due no doubt in part to the rather complicated tensor version (\ref{WLt})
of its relationship to the Weyl tensor, and also to various
mistakes in some papers, the Lanczos potential has failed to attract major
attention, and there is perhaps still an air of uncertainty surrounding it.

Although Bampi and Caviglia \cite{BC} identified the flaw in Lanczos'
original attemt to prove its existence, the complicated nature of their
own existence proof also helped to set the Lanczos potential apart.

Although Maher and Zund \cite{MZ} had discovered the very simple and natural
spinor structure of $L_{ABCA'}$ as early as 1968, this result attracted
little interest, perhaps because of some mistakes and misprints in this
and subsequent papers of Zund's.

Twenty years later, Illge's work \cite{Illge} highlighted and exploited the
spinor representation, and also discovered for the first time the remarkably
simple wave equation for the Lanczos potential of the Weyl spinor in vacuum
spacetimes  and Lanczos differential gauge. (Although Lanczos had calculated a
wave equation for the Lanczos potential of the Weyl tensor in tensor notation,
containing complicated non-linear terms obtained by everywhere replacing
$C_{abcd}$ with the appropriate expression in $L_{abc}$, it contained some
mistakes, which were repeated, or only partly corrected by others; no-one had
suspected that these non-linear terms were actually identically zero in four
dimensions.) The relative simplicity of the Lanczos spinor wave equation in
the less ideal cases of non-vacuum, arbitrary differential gauge, arbitrary 
$W_{ABCD}$ (in particular it is linear) enabled Illge to use the wave 
equation in his somewhat indirect proof of existence of the Lanczos potential.
More precisely he showed the equivalence of the two solution sets of the 
wave equation, subject to an initial value constraint and the Weyl-Lanczos
equation. On the otherhand, Illge has established {\it uniqueness} results
as well as existence, in his proof.

In this paper we have given an alternative very direct proof of existence in
Section 3; the essential step in our proof of existence is simply appealing to
the wave equation. We hope this simple proof, and the direct link with the
familiar wave equation for a Hertz-like potential will highlight unambiguously
the very natural and familiar structure of the Lanczos potential for the Weyl
spinor, and open up the way for deeper considerations.

Further investigation is also needed to decide whether the Hertz-like potential
$T_{ABCD}$ has more significance; certainly it is useful in obtaining, for the
first time,  an explicit expression for the gauge freedom in $L_{ABCA'}$, in
Section 4.

By applying the spinor results of Sections 3, 4, 5 to electromagnetic theory in
Section 6 we emphasized, as pointed out by Illge \cite{Illge} that the existence
of the electromagnetic potential is not dependent on the second of Maxwell's
equations, via Poincare's lemma, which is the way in which it is usually
presented. In electromagnetic theory when the electromagnetic potential
$A_{AA'}$ is hermitian, the two superpotentials $T_{AB}$ and $H_{A'B'}$ are
essentially equivalent, and in fact are seen to be a spinor version of a
Hertz-like potential; of course  such a direct relationship is not possible for
the two superpotentials of $W_{ABCD}$. Also in electromagnetic theory, as
mentioned above the potential $A_{AA'}$ is hermitian; this simplification cannot
apply to $L_{ABCA'}$ either; however, such a possibility exists for the
potential $H_{ABA'B'}$ of $L_{ABCA'}$ (for, at least, a significant class of
spacetimes), and this is one of the questions requiring further investigations.

The existence of a potential such as $L_{ABCA'}$ for $\Psi_{ABCD}$ is of course
well known and thoroughly investigated in flat space in connection with the
massless field equation; and indeed  a chain  of Hertz-like potentials,
including some analogous to $T_{ABCD}$ and $H_{ABA'B'}$, have been studied.
Although Penrose \cite{PR2} has studied these using spinor techniques, his
results are strictly for (conformally) flat spaces.  In ${\cal H}$-spaces
(complex general relativity) the complex connection plays the role of a complex
Lanczos potential $L_{ABCA'}$ of one of the Weyl spinors (recall that the other
Weyl spinor is zero since ${\cal H}$-spaces are always left-flat), and this
potential itself {\it always} permits a potential $H_{ABA'B'}$. This
$H$-potential is the basis for constructing physics in ${\cal H}$-spaces.
This is part of our motivation for investigating, in Section 5, the existence
of an $H$-potential in real curved space. It is hoped, having now shown that
such a potential does exist in physically important curved spaces, that
(at least part of) the successful programme associated with the complex
$H$-potential can be applied to this $H$-potential in real spacetimes.

A related motivation is that in earlier investigations of Lanczos
potentials, the existence of such an $H_{ABA'B'}$ was not only an
important aid to calculate the Lanczos potential $L_{ABCA'}$, but
the possibility of it having physical and geometrical significance
has also been considered. We summarize those cases below:

\begin{itemize}
 \item Torres del Castillo \cite{TdC1} has studied spacetimes admitting a
       normalized spinor dyad $(o^{A},\iota^{A})$ in which
       $$
        \kappa=\sigma=0
       $$
       and in which the Ricci spinor satisfies
       $$
        \Phi_{ABA'B'}o^{A}o^{B}=0
       $$
       He found that in all such spaces there exists a Lanczos potential
       $L_{ABCA'}$ of the Weyl spinor such that $L_{ABCA'}$ can be written
       $$
        L_{ABCA'}=\nabla_{(A}{}^{B'}H_{BC)A'B'}
       $$
       for some completely symmetric spinor $H_{ABA'B'}$. By defining
       $$
        \eta_{ab}=g_{ab}-H_{ab}
       $$
       where $H_{ab}$ is the symmetric, trace-free tensor equivalent of 
       $H_{ABA'B'}$ Torres del Castillo obtained a complex, conformally flat 
       metric $\eta_{ab}$.
 
 \item Bergqvist and Ludvigsen \cite{BL} define a flat connection in 
       the Kerr spacetime, by
       $$
        \hat{\nabla}_{AA'}\xi^{B}=\nabla_{AA'}\xi^{B}+2\Gamma_{C}{}^{B}
        {}_{AA'}\xi^{C}
       $$
       where
       \begin{equation}
        \Gamma_{ABCA^{\prime}}=\nabla_{(A}{}^{B^{\prime}}
        H_{B)CA^{\prime}B^{\prime}}
        \label{}
       \end{equation}
       and $H_{ABA'B'}$ is hermitian and given by
       \begin{equation}
        H_{ABA^{\prime}B^{\prime}}=\frac{\rho+\bar{\rho}}{4\rho^{3}}\Psi_{2}
        o_{A}o_{B}o_{A^{\prime}}o_{B^{\prime}}.
        \label{}
       \end{equation}
       where $o^{A}$ is a principal spinor of the Weyl spinor. Subsequently
       Bergqvist \cite{Bergqvist} has shown that $\Gamma_{(ABC)A^{\prime}}$
       is a Lanczos potential in the Kerr spacetime. This connection has been
       used by Bergqvist and Ludvigsen \cite{BL} to construct quasi-local
       momentum in the Kerr spacetime. 

 \item In \cite{AE} these results are generalized to Kerr-Schild spacetimes
       i.e.,
       $$
        g_{ab}=\eta_{ab}+fl_{a}l_{b}
       $$
       where $\eta_{ab}$ is a flat metric, $l^{a}$ is null and $f$ is 
       a real function. It is shown that providing $l^{a}$ is geodesic and
       shear-free (or if another more technical condition is fulfilled)
       then $H_{ab}=fl_{a}l_{b}$ is a hermitian $H$-potential of a 
       Lanczos potential of the Weyl spinor, that also defines a 
       curvature-free connection (See also \cite{Harnett}.).
        
 \item In a recent paper \cite{LMO} L\'opez-Bonilla et. al. have found, for
       the Kerr spacetime, an explicit Lanczos potential of the Weyl spinor,
       given by a hermitian $H$-potential of the type discussed in this paper,
       for the Kerr spacetime.

 \item Novello and Velloso \cite{NV} have shown that for perfect fluid
       spacetimes that admit a normalized timelike vector field $u^{a}$,
       $u_{a}u^{a}=1$ which is shear-free and vorticity-free, so that
       $$
        \nabla_{a}u_{b}=u_{a}\dot{u}_{b}+\frac{1}{3}\theta h_{ab}
       $$
       where $\dot{u}_{a}=u^{b}\nabla_{b}u_{a}$, $h_{ab}=g_{ab}-u_{a}u_{b}$ 
       and $\theta$ is the expansion of $u^{a}$, then the tensor
       \begin{equation}
        L_{abc}=2\dot{u}_{[a}u_{b]}u_{c}+\frac{2}{3}g_{c[a}\dot{u}_{b]}
        \label{NVLanczos}
       \end{equation}
       is a Lanczos potential of the Weyl spinor (the second term is to ensure
       that $L_{ab}{}^{b}=0$). It is easy to confirm that when
       \begin{equation}
        H_{ab}=u_{a}u_{b}-\frac{1}{4}g_{ab}=\frac{3}{4}u_{a}u_{b}-
        \frac{1}{4}h_{ab}. \label{NVH}
       \end{equation}
       is substituted for $H^{1}$ (with $H^{2}=0$) into equation (\ref{H1H2})
       we obtain precisely the Lanczos potential (\ref{NVLanczos}). 
\end{itemize}

We conclude with two comments. In some of the examples quoted above an
$H$-potential of a Lanczos potential of the Weyl spinor was found for some
non-Einstein spacetimes; it remains an open question if such a construction
is possible for a significant class of non-Einstein spacetimes. Earlier in
this section we commented on the possible significance of having a hermitian
$H$-potential. We note, from our examples above, that in the cases where this
potential was used for constructing curvature-free connections and  quasi-local
momentum, it was hermitian; therefore, it would appear that if these
constructions are to be possible in other spaces, we need to know if hermitian
superpotentials for the Weyl spinor, can be found for other spacetimes.

\section{Acknowledgements}

SBE is grateful to NFR (the Swedish Natural Science Research Council) 
for financial support.

\end{document}